\documentclass[aps,twocolumn,superscriptaddress,floatfix]{revtex4-1}

\usepackage{lipsum}
\usepackage{graphicx}
\usepackage{amsmath,amssymb}
\usepackage{braket}
\usepackage{mathtools}
\usepackage{hyperref}
\usepackage{multirow}
\usepackage{color}
\usepackage[normalem]{ulem}
\usepackage{amsfonts}
\usepackage{float}
\usepackage{pdfpages} 
\makeatletter
\usepackage{subfigure}
\usepackage{soul, xcolor}

\setstcolor{red}

\usepackage{tikz}

\def\bra#1{\mathinner{\langle{#1}|}}
\def\ket#1{\mathinner{|{#1}\rangle}}
\def\braket#1{\mathinner{\langle{#1}\rangle}}

\def\beq{\begin{equation}}
\def\eeq{\end{equation}}
\def\bea{\begin{eqnarray}}
\def\eea{\end{eqnarray}}

\usepackage{pdfpages} 
\makeatletter
\AtBeginDocument{\let\LS@rot\@undefined}
\makeatother

\begin{document}

\title{Subdiffusive transport in the Fredkin dynamical universality class}

 \author{Catherine McCarthy}
\affiliation{Department of Physics, University of Massachusetts, Amherst, MA 01003, USA}

 \author{Hansveer Singh}
\affiliation{Department of Physics, University of Massachusetts, Amherst, MA 01003, USA}
\affiliation{Max Planck Institute for the Physics of Complex Systems, 01187 Dresden, Germany}

\author{Sarang Gopalakrishnan}
\affiliation{Department of Electrical and Computer Engineering, Princeton University, Princeton, NJ 08544, USA}

\author{Romain Vasseur}
\affiliation{Department of Physics, University of Massachusetts, Amherst, MA 01003, USA}
\affiliation{Department of Theoretical Physics, University of Geneva, 24 quai Ernest-Ansermet, 1211 Geneva, Switzerland}

\begin{abstract}

Fredkin-type kinetic constraints appear in systems ranging from classical interfaces to the Fredkin and Motzkin quantum spin chains.  Here, we identify a  pseudolocal conserved charge associated with Fredkin-constrained dynamics that may be leveraged to subdiffusively bound the dynamical exponent $z \geq 5/2$. Our results establish that transport in Fredkin-constrained systems belongs to a distinct dynamical universality class that is separate from ordinary diffusion.  This universality class is stable to local perturbations that conserve the pseudolocal charge. 

\end{abstract}
\vspace{1cm}

\maketitle

%



\section{Introduction}

Hydrodynamics is an effective field theory description of the transport properties of locally-conserved densities in many-body systems \cite{ghdperspective, DeNardis_review2022, eftlectures, doyon_ghdlecture, RevModPhys.93.025003}.  The central pillar of hydrodynamics is that, absent any special symmetries, locally-conserved densities relax rapidly; this relaxation process typically yields diffusive transport of conserved charges.  The insensitivity of hydrodynamics to the microscopic details of a system gives rise to a universal description for transport in chaotic systems.  Consequently, transport in both classical and quantum systems at nonzero temperature is generically expected to be diffusive.


Given the universality of ordinary diffusion, finding exceptions to diffusive behavior has been a longstanding quest.  One of the most famous exceptions to diffusion is the Kardar-Parisi-Zhang universality class, which predicts superdiffusive interface growth in a wide variety of stochastic systems \cite{KPZ}.  In the search to identify new dynamical universality classes, one promising approach is the study of kinetically-constrained models.  In stochastic systems with kinetic constraints, which were first introduced as toy models of the glass transition~\cite{ritort2003glassy}, certain local rearrangements of conserved charges are forbidden.  Fracton hydrodynamics in 1D systems has recently emerged as an important example of subdiffusive, kinetically-constrained dynamics that belongs to a distinct dynamical universality class separate from diffusion~\cite{FractonHydro, PhysRevLett.125.245303, FractonSubdiffusion, glorioso2021breakdown, PabloFragmentPRX, morningstarKineticallyConstrainedFreezing2020}.  The kinetic constraints in fractonic systems are the result of imposing charge and multipole conservation laws-- for systems that conserve the $n^{\rm th}$ multipole moment, dynamical spacetime scaling $x \sim t^{1/z}$ is characterized by the dynamical exponent $z=2n$.  Nonlinear effects may be also be incorporated into fracton hydrodynamics to give transport with certain fractional power laws~\cite{glorioso2021breakdown, guo2022fracton}.  To date, fracton hydrodynamics is the only well-established dynamical universality class apart from diffusion known to arise from a kinetically-constrained model.


Another potentially-distinct dynamical universality class results from imposing what we will refer to as the ``Fredkin'' constraint in one-dimensional systems.  The Fredkin constraint is defined in terms of mapping local degrees of freedom to a height field; for Fredkin-constrained dynamics, local charge rearrangements are forbidden from locally decreasing the minimum value of the height field.  Fredkin-constrained dynamics arise in contexts ranging from interface growth models \cite{Koduvely1998} and deposition-evaporation models \cite{tde_orig, tde_numerics} to entanglement transitions in random unitary circuits \cite{Morral2024} but have been studied most extensively in the context of the Fredkin and Motzkin quantum spin chains \cite{PhysRevLett.109.207202, FredkinEntanglement, Korepin2016fredkin, movassagh2016gap, ChenFredkin2017, ChenFredkin2017a, hans_subdiffusion, hans_staircase, PhysRevResearch.4.L012003, PhysRevE.106.014128}.  The dynamical exponent for Fredkin-constrained dynamics has previously been diffusively lower-bounded ($z \geq 2$)~\cite{movassagh2016gap}. Numerical evidence suggests that this bound is not tight and that transport is subdiffusive at half-filling, with reported values of the dynamical exponent ranging from $z=5/2$ to $z=3.16$~\cite{PhysRevResearch.4.L012003, ChenFredkin2017, ChenFredkin2017a, hans_subdiffusion, Koduvely1998,PhysRevB.104.115149}.  The precise value of the dynamical exponent is an open question, and the mechanism underpinning any $z > 2$ remains unclear-- the fractional value of $z$ is incompatible with simple linear hydrodynamics involving strictly local charges, of the type considered in Ref.~\cite{FractonHydro}. Possible mechanisms for this exponent include intrinsically nonlinear mechanisms~\cite{popkov2015fibonacci} or pseudolocal conserved charges with power-law tails~\cite{BOUCHAUD1990127}.    More fundamentally, there is currently no proof that Fredkin-constrained dynamics have $z > 2$, and it remains controversial whether subdiffusion truly persists to arbitrarily late times in the thermodynamic limit.

In this work, we present an argument that rigorously lower-bounds the dynamical exponent $z \geq 5/2$ for Fredkin-constrained dynamics at half-filling, definitively proving that Fredkin-constrained dynamics is subdiffusive and constitutes its own dynamical universality class separate from diffusion.  To do so, we identify a pseudolocal conserved charge-- which we will refer to as the ``matched parentheses charge''-- in 1D Fredkin-constrained Markov processes and explore the consequences of its presence.  We demonstrate that this charge possesses algebraically-decaying tails that underpin all nontrivial static correlations relevant to Fredkin-constrained systems.  Despite its pseudolocality, we prove that the matched parentheses charge is both extensive and additive, thus making it relevant to transport.  Taking advantage of the correspondence between the scaling of the gap of frustration-free Hamiltonians and the dynamics of classical Markov processes, we use the Hamiltonian of the Fredkin quantum spin chain to construct a variational spin-wave ansatz in terms of the new charge.  This ansatz produces an exact upper bound for the gap of the Hamiltonian  $\Delta \leq  {\cal O}(L^{-5/2})$ and consequently a lower bound for the dynamical exponent, $z \geq 5/2$; with minor modifications, this result may also be exactly shown for Motzkin spin chains (see Appendix~\ref{app:motzkin}).  The subdiffusive $z \geq 5/2$ bound establishes that Fredkin-constrained dynamics belongs to its own, distinct dynamical universality class.

\section{Background}

\subsection{The Fredkin constraint}

The most general formulation of the Fredkin constraint involves mapping a system's local degrees of freedom to a height field.  For example, a binary system in one dimension may be interpreted as a 1D lattice consisting of particles and holes.  Mapping this system to a height field requires starting at some point $x=0$ where $h(0)=0$; the height field at any other point is given by $h(x) = \sum_{n<x} \Delta h_n$, where $\Delta h_n= 1$ for a particle and $\Delta h_n = -1$ for a hole (fig. \ref{fig:mapping}a).  The Fredkin constraint forbids any rearrangement of local degrees of freedom that decreases the local minimum value of the height field. In the particle/ hole language, the height field constraint can be interpreted as a 4-site ``controlled-swap'' restriction on local rearrangements of charges (fig.~\ref{fig:mapping}b) although in principle, the height-field constraint may be implemented over an arbitrary number of sites.

\subsection{Mapping between Markov processes and quantum spin chains}

While Fredkin-constrained dynamics are generic and may appear in many contexts, the derivation of the dynamical exponent in this work will make use of a correspondence between the dynamics of Markov processes and the gap scaling of critical frustration-free Hamiltonians at a Rokshar-Kivelson (RK) point~\cite{U1FRUC, SanjayAmosFracton, hans_subdiffusion}. 

Consider a classical discrete-time Markov process described by a probability density $\vec{p}(t)$ over all possible states.  At time $t=0$, let $p_\alpha(0) = 1$ and all $p_{\beta \neq \alpha}(0)=0$.  Under stochastic time evolution, the probability density will relax to an equilibrium state in which all possible states are assigned equal probabilities.  The mixing time associated with this relaxation process can be expressed as the inverse of the ``gap'' of this Markov process, $\Delta_M$; if the Markov process is gapless, we expect $\Delta \rightarrow 0$ in the thermodynamic limit.

The discrete-time Markov process can be viewed as a Trotterized version of a stochastic continuous-time process $\vec{p}(t+\tau) = T \vec{p}(t)$, where $T$ is the classical transfer matrix describing the Markov process. In the continuous time limit $\tau \to 0$, the transfer matrix $T$ may be related to a quantum Hamiltonian $H$ through the usual relation $T \simeq 1 - \tau H$.  The class of quantum Hamiltonians that are related to Markov processes through standard quantum-classical mappings are known as \textit{Rokhsar-Kivelson} (RK, or frustration-free) Hamiltonians. The gap of the Markov matrix $\Delta_M$ is then directly related to the gap $\Delta$ of $H$, and therefore, the scaling of $\Delta_M$ is the same as the gap-scaling of the RK Hamiltonian $\Delta_M \sim \Delta \sim L^{-z}$ \cite{SanjayAmosFracton}.

\subsection{The Fredkin spin chain}

To make use of the mapping described in the previous section, in this work we focus on a pair of one-dimensional quantum spin chains that map to Fredkin-constrained Markov processes: the Fredkin (spin-$\frac{1}{2}$) ~\cite{Korepin2016fredkin} and Motzkin (spin-1)\cite{PhysRevLett.109.207202, FredkinEntanglement} chains.  Both spin chain variants have frustration-free Hamiltonians with an exactly-known critical ground state.  The energy gap of these Hamiltonians are rigorously upper-bounded as $L^{-2}$ for a system of size $L$~\cite{movassagh2016gap} ---therefore, despite being a one-dimensional critical point, they are not described by a conformal field theory~\cite{ChenFredkin2017,ChenFredkin2017a}. 

For brevity, we limit the discussion in the main text to the Fredkin spin chain, and all results are reproduced for the Motzkin spin chain in Appendix~\ref{app:motzkin}.  The Fredkin Hamiltonian is given by 
\begin{equation}
H = \frac{1}{8}\sum_i (1+\sigma^z_i)(1-\vec{\sigma}_{i+1}\cdot\vec{\sigma}_{i+2})+(1-\sigma^z_{i+2}) (1-\vec{\sigma}_{i}\cdot\vec{\sigma}_{i+1}),
\label{equation:Hproj}
\end{equation}
where $\sigma^{\alpha}_i$ are the usual Pauli operators with $\alpha=x,y,z$. The Fredkin Hamiltonian conserves the total magnetization, $S^z_{\text{tot}} = \frac{1}{2}\sum_i \sigma^z_i$ and exhibits charge-parity (${\rm CP}$) symmetry. Unless otherwise specified, we will consider periodic boundary conditions and restrict to the $S^z_{\text{tot}}=0$ sector where the ground state is a uniform superposition of all $S^z_{\text{tot}}=0$ states.   Following the usual quantum-to-classical correspondence combined with the frustration-free property, the Fredkin Hamiltonian maps onto a classical Markov process of constrained hard-core random walkers~\cite{hans_subdiffusion}, where an up spin corresponds to a particle, and down spin to a hole. 

\begin{figure*}[t!]
    \centering
    \includegraphics[width=0.99\textwidth]
    {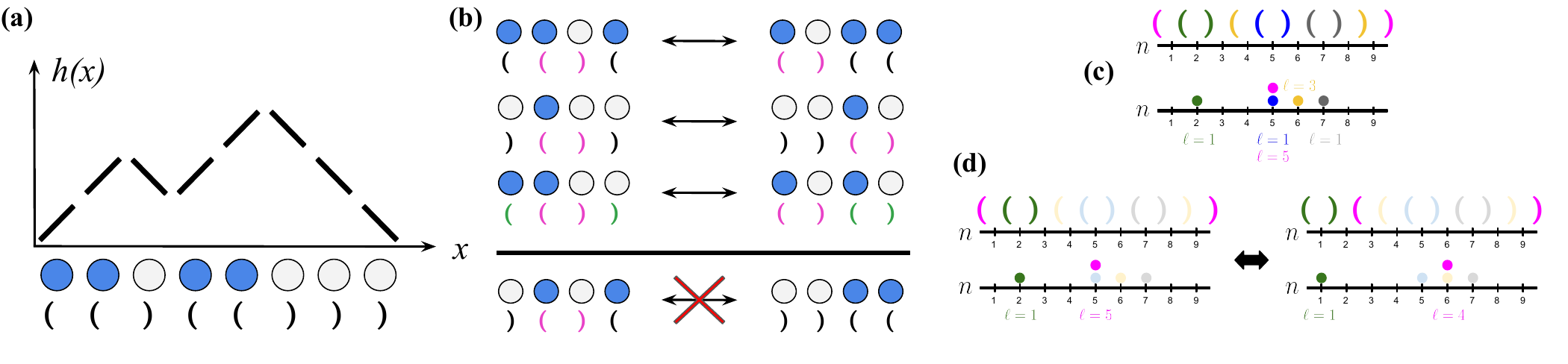}
    \caption{\textbf{The matched parentheses charge and the Fredkin constraint} (a) Three representations (height field, random walkers, and Dyck alphabet) of a classical Markov process in one dimension with two degrees of freedom.  (b)  The Fredkin constraint prohibits rearrangements of local degrees of freedom that decrease the local minimum value of the height field.  In each configuration shown, the particle may hop between the second and third sites if the neighboring sites satisfy the height-field constraint; the three moves that satisfy the height field constraint are shown on the top, while the disallowed move that changes the local minimum of the height field is shown at the bottom.  In the Dyck alphabet representation, the allowed moves locally conserve the number (but not the type) of irreducible Dyck words, while the disallowed move destroys a Dyck word. (c) A configuration of parentheses (top) translated into a configuration of matched parentheses charges (bottom).  Each matched parentheses charge sits on on the central link $n$ between the two matched parentheses and has semilength $\ell$ that is half of the distance between the matched parentheses.  (d) An example of a move allowed under the Fredkin Markov process, in which the second and third parentheses in the configuration swap positions.  Allowed Fredkin moves conserve the number of matched parentheses charges and correspond to the local hopping of these charges.}
    \label{fig:mapping}
\end{figure*}

\subsection{Dyck words}

As originally outlined in~\cite{Korepin2016fredkin}, the Fredkin Hamiltonian naturally lends itself to a combinatorial description in terms of the Dyck alphabet.  We briefly summarize some important aspects of the Dyck alphabet here.  The Dyck alphabet consists of two symbols: open and closed parentheses.  Dyck words are defined as strings of parentheses in which every parenthesis is matched when read from left to right; for instance, (()) and ()() are both Dyck words, while )()( is not.  A Dyck word necessarily contains an even number of parentheses and can be identified by its semilength, $\ell$, which is half of the total number of parentheses.  The number of Dyck words of semilength $\ell$ is given by the $\ell^{\rm th}$ Catalan number, $C_\ell$, which scales asymptotically as $C_\ell \sim \ell^{3/2}$. A class of Dyck words that will be important for the following discussion are irreducible Dyck words. Irreducible Dyck words are those in which the first parenthesis in the word is matched by the final parenthesis in the word; for example, ((())) and (()()) are both irreducible Dyck words of semilength $\ell=3$, while (())(), ()(()), and ()()() are also Dyck words of semilength $\ell=3$ but are not irreducible.    Since irreducible Dyck words of semilength $\ell$ are of the form (...), where ... must itself be a Dyck word, there are $C_{\ell-1}$ irreducible words of semilength $\ell$.

\section{The matched parentheses charge}

By assigning one type of parenthesis to each degree of freedom, a Markov process with two degrees of freedom may be expressed in terms of the Dyck alphabet(fig.~\ref{fig:mapping}).  The Fredkin constraint is defined in terms of prohibiting moves that decrease the local minimum of the height field.  In the language of the Dyck alphabet, the conservation of the local minimum of the height field translates into the local conservation of irreducible Dyck words, as illustrated in fig.~\ref{fig:mapping}b.  Since this charge reflects the conservation of the number of matched parentheses in the system, we will refer to irreducible Dyck words as ``matched parentheses charges.'' 

To illustrate the conservation law, consider the configuration $(()()())$.  This configuration consists of four irreducible Dyck words, three of semilength $\ell=1$ and one of semilength $\ell=4$.  We define the position of each irreducible Dyck word by the location of its midpoint; in this configuration, the $\ell=1$ words are located on links $n=2,4,$ and $6$, while the midpoint of the $\ell=4$ word falls on link $n=4$.  Consider swapping the second and third sites in the above configuration, which is permitted under Fredkin-constrained dynamics.  The word of semilength $\ell=4$ and the $\ell=1$ word originally located on link $n=2$ are affected by this move, while the other two $\ell=1$ words are unaffected.  This process is illustrated below, with the affected words underlined with their midpoints marked with a vertical bar:
\begin{equation}
    \underline{(\underline{(|)}(|)())} \rightarrow \underline{(|)}~\underline{(()|())}.
    \label{eq:allowed}
\end{equation}

After this allowed  move, the word that was originally located on link $n=2$ with semilength $\ell=1$ has moved to link $n=1$, while the word that originally had semilength $\ell=4$ whose midpoint was located on link $n=4$ has moved to link $n=5$ and has decreased in semilength to $\ell=3$.  

This example captures a general characteristic of moves permitted under Fredkin constraints: all allowed moves can be interpreted as a Dyck word of semilength $\ell=1$ moving past another parenthesis (see fig. 1\ref{fig:mapping}b).  In the $S_z^{\rm tot}=0$ charge sector with periodic boundary conditions, every parenthesis is the endpoint of a Dyck word.  The effect of moving an $\ell=1$ Dyck word past the endpoint of some other word of semilength $\ell$ is to move that word's midpoint by a single link to $n \pm 1$ while changing its semilength by one to $\ell \mp 1$.

It is also instructive to consider a move that is forbidden under Fredkin-constrained dynamics:
\begin{equation}
    (()()()) \not\rightarrow (())(()).
    \label{eq:forbidden}
\end{equation}

The initial configuration has two irreducible Dyck words of semilengths $\ell=1$ and $\ell=4$ whose midpoints fall on link $n=4$.  The final configuration corresponds to the destruction of both of these Dyck words and the creation of two $\ell=2$ Dyck words whose midpoints fall on links $n=2$ and $n=6$.  In contrast to the process depicted in \ref{eq:allowed}, the process shown in \ref{eq:forbidden} does not involve the movement of an $\ell=1$ Dyck word past the endpoint of some other Dyck word.  Therefore, despite the fact that \textit{globally}, the total number of irreducible Dyck words remains unchanged, the process illustrated in \ref{eq:forbidden} violates the \textit{local} conservation of irreducible Dyck words.

In summary, all allowed Fredkin moves can be interpreted as the local hopping of irreducible Dyck words, while moves forbidden under Fredkin-constrained dynamics correspond to the local destruction (and creation) of Dyck words.  The midpoint of an irreducible Dyck word can then be interpreted as a conserved charge-- we accordingly call this the ``matched parentheses charge.''  Each matched parentheses charge has two labels-- its position, $n$, and its semilength, $\ell$.  In the $S_z^{\rm tot}$ charge sector with periodic boundary conditions, the hopping of an $\ell=1$ charge is always accompanied by the movement of another charge-- if this charge was originally of semilength $\ell$ and located on link $n$, then moving the $\ell=1$ charge past one of its endpoints will cause it to hop to link $n \pm 1$ and change in semilength to $\ell \pm 1$.

The matched parentheses charge is also conserved by the Hamiltonian of the Fredkin quantum spin chain.  The Fredkin Hamiltonian may be re-written so as to make the presence of the matched parentheses charge more apparent.  First, we introduce the hard-core bosonic creation and annihilation operators of the matched parentheses charge, $(a_{i}^{\ell})^{(\dagger)}$. The number operator may be written as $q_{i}^{\ell} = (a_{i}^{\ell})^\dag a_{i}^{\ell}$.  In the discussion that follows, we adopt the convention that the lower index denotes position and the upper index denotes semilength; if either index is irrelevant to the discussion at hand, it will be omitted and assumed to be summed over.

Within the $S_z=0$ charge sector with periodic boundary conditions, the Hamiltonian may then be expressed as:
\begin{equation}
\begin{split}
    &H  = \frac{1}{2}\sum_{i,\ell}q_{i-1}^{\ell+1} q_{i-\ell}^{1} + q_{i}^{\ell}q_{i-\ell-1}^{1} + q_{i+1}^{\ell+1}q_{i+\ell}^{1} + q_{i}^{\ell}q_{i+\ell+1}^{1}\\
    &-(a_{i-1}^{\ell+1})^\dag (a_{i-\ell}^{1})^\dag  a_{i}^{\ell} a_{i-\ell-1}^{1}-(a_{i+1}^{\ell+1})^\dag (a_{i+\ell}^{1})^\dag a_{i}^{\ell} a_{i+\ell+1}^{1}-\text{h.c.},
\end{split}
\end{equation}

In this representation one can see that the total matched parentheses charge is $Q_{\rm tot} = \sum_{i,\ell}q_{i}^{\ell}$.

\section{Pseudolocality of the matched parentheses charge}


The total matched parentheses charge $Q_{\rm tot} = \sum_{n} q_n$ is related to $S^z_{\rm tot}$; for example, $Q_{\rm tot}=L/2$ for a system of $L$ spins with $S^z_{\rm tot}=0$ with periodic boundary conditions. As a result, any spin-$\frac{1}{2}$ Hamiltonian that preserves $S^z_{\rm tot}$ acting on a system with periodic boundary conditions will globally conserve the number of matched parentheses charges.    In contrast, the Fredkin Hamiltonian has a stronger, local conservation law governing the matched parentheses charge and so one has a corresponding continuity equation (see Appendix~\ref{section:additivity}).  The locality of the matched parentheses charge is not entirely obvious \textit{a priori}-- the total matched parentheses charge may be written as the sum of terms $Q_{\rm tot} = \sum_n q_n$, but the support of each charge is not obviously bounded. Here, we first show that the matched parentheses charge has power-law tails that account for nontrivial static correlations between parentheses. We then prove that the charge is additive and extensive, thus demonstrating that even with its power-law tails, the charge is relevant to transport.




First, we consider the source of the algebraically-decaying tails of the matched parentheses charge.  All static correlations between matched parentheses charges arise from the fact that a parenthesis may only function as the endpoint of a single charge of semilength $\ell$.  To illustrate this, consider a charge corresponding to a configuration $()$ at the origin, $n=0$ and $\ell=1$.  Since parentheses may not be shared between charges, and since the existence of a charge of semilength $\ell'=1$ at site $m=\pm1$ charge would require one of the endpoints of the $n=0, \ell=1$ charge to instead be one of the endpoints of the $m=\pm 1, \ell'=1$ charge, the simultaneous existence of both charges is impossible.  Extending this logic to an arbitrary link $\ell'$, the presence of a charge with $n=0, \ell=1$ precludes the existence of any $m=\pm \ell', \ell'$ charge.   The tails of the matched parentheses charge therefore depend on the probability of finding a charge of semilength $\ell'$ that lives $\ell'$ links away from a charge of semilength $\ell=1$.  Since the probability of finding a charge of semilength $\ell$ on an arbitrary site is given by $C_{\ell-1}/2^{2 \ell} \sim \ell^{-3/2}$, all nontrivial static correlations between the presence of a charge at the origin with semilength $\ell=1$ and any other charges are given solely by the probability of finding $q_{\pm \ell'}^{\ell'}$.  Therefore, we find $\langle q_m^1 q_n \rangle \sim |m-n|^{-3/2}$.

For a charge to be relevant to thermodynamics, it must be extensive and additive; despite having power-law tails, the matched parentheses charge indeed exhibits both of these properties. To see this, consider a region of $2L \rightarrow \infty$ total sites split into two $L$-site subsystems $A$ and $B$. The total charge in this region $Q_{AB} = \sum_{n=1}^{2L} q_n$. The total number of matched parentheses charges in subsystem $A$, given by $Q_A = \sum_{n=1}^{L} q_n$, grows with $L$, and so $\langle Q_A \rangle \sim L$.  One could worry that in spite of the extensivity of the matched parentheses charge, the power-law tails of the charge may lead to $Q_{AB} \neq Q_A + Q_B$.  From showing exactly that $\langle Q_A^2 \rangle_c \sim L$, we conclude that fluctuations of the total charge are sufficiently suppressed to guarantee additivity $Q_{AB} \simeq Q_A + Q_B$, up to subleading corrections of $\mathcal{O}(\sqrt{L})$ -- for further details, see Appendix~\ref{section:additivity}.   Therefore, fluctuations in the value of $Q_A$ are suppressed as $\sqrt{\langle Q_A^2 \rangle_c}/{\langle Q_A \rangle} \sim 1/\sqrt{L}$, so the local density of matched parentheses is well-defined.  Since the matched parentheses charge is locally conserved, extensive, and additive, it is relevant to hydrodynamics.  Intuitively, due to the fact that each matched parentheses charge is tied to two individual parentheses, both of which move with $z \approx 5/2$, the matched parentheses charge should have the same dynamical exponent as the original charge-- this intuition can be confirmed numerically (see Appendix~\ref{section:numerics}).

\begin{figure}[t!]
    \centering
    \includegraphics[width=0.99\columnwidth]
    {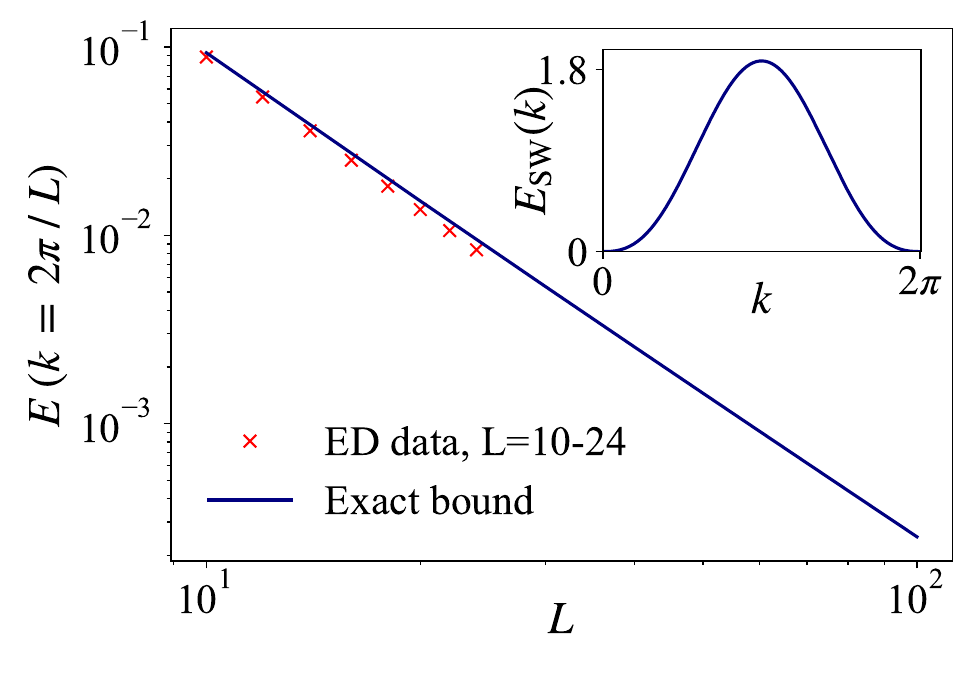}
    \caption{\textbf{Exact bound on gap scaling.}  Exact diagonalization (ED) data shown for system sizes of $L=10-24$ sites plotted with the exact $\sim L^{-5/2}$ analytical bound.  Inset: exact expression for the dispersion $E_{\mathrm{SW}}(k)$ of the spin-wave ansatz, evaluated in the limit $L \rightarrow \infty$.}
    \label{fig:sw}
\end{figure}

\section{Exact subdiffusive bound on the dynamical exponent}

The gap of the Fredkin Hamiltonian scales polynomially with system size $L$ as $\Delta \sim L^{-z}$~\cite{movassagh2016gap}.  To produce a bound for the gap, and by extension the dynamical exponent, we construct a variational spin-wave ansatz in terms of the matched parentheses charge -- see Ref.~\cite{PhysRevLett.131.220403} for a similar approach applied to other systems with fracton-like constraints.  We find that $z \geq z_{\rm SW} = 5/2$ and derive an exact expression for this bound in Appendix~\ref{section:bound}; below, we summarize the procedure to do so.

We construct a spin-wave trial wave function in terms of the matched parentheses charge $q$ for the state $\ket{k} = \sum_n e^{i k \cdot n} q_n \ket{GS}$ with momentum $k = 2\pi j/L$, for $j=0, 1, ..., L-1$, where the $S^z_{\rm tot}=0$ ground state $\ket{GS}$ is given by the uniform superposition of all $S^z_{\rm tot}=0$ states.  By standard variational considerations, the energy of the spin-wave state $\ket{k}$, given by $E_{\rm SW}(k) = \braket{k|H|k}/\braket{k|k}$, serves as an upper bound on the ground state energy of the Fredkin Hamiltonian within the $k^{\rm th}$ crystal momentum sector.  The gap of the Fredkin Hamiltonian may then be bounded by evaluating the energy of the spin-wave state in the $k=2\pi/L$ sector.



Using translation invariance, we have $\braket{k|H|k} = L \sum_r e^{ik\cdot n} \braket{GS| q_0 H q_n | GS}$, and using  $\braket{k|k}\sim L$ (see Appendix~\ref{section:bound}), the task of bounding the gap reduces to evaluating the Fourier transform of the correlator $C(n)\equiv \braket{GS|q_0 H q_n|GS}$.  From $H\ket{GS}=0$, we may write $C(n) = \braket{GS|q_0 [H, q_n|GS}$ with $i[H, q_n] = j_n - j_{n+1}$, where $j_n$ is the current of the matched parentheses charge on site $n$. This yields
\begin{equation}
C(n) = i\braket{GS|q_0 j_{n+1}|GS}-i\braket{GS|q_0 j_{n}|GS}.
\end{equation}
We obtain exact expressions for the charge-current correlator $\braket{GS|q_0 j_n|GS}$ in the $S^z_{\rm tot}=0$ ensemble for any system size $L$; for further details, see Appendix~\ref{section:bound}. We find that $\sum_n n^2 C(n) =0$, ruling out diffusive $\sim k^2$ contributions to $E_{\rm SW}(k)$. Instead, the low momentum scaling
of $E_{\rm SW}(k) \sim k^{5/2}$ is governed by the long distance decay of $C(n)  \sim n^{-7/2}$, which follows from the charge-current correlator $\braket{GS|q_0 j_{n+1}|GS}$ decaying as $\sim n^{-5/2}$.  This $\sim n^{-5/2}$ decay may be readily obtained by computing the probability that processes such as $(\dots ()) \to (\dots )() $ occur, since these processes make up all possible contributions to the charge-current correlator (Appendix~\ref{section:bound}). 

Our results provide the exact energy of the spin-wave state $E_{\rm SW}(k)$ for any system size $L$, which we verify constitutes an upper bound of the gap of the Fredkin Hamiltonian (Fig.~\ref{fig:sw}). In the thermodynamic limit, we find the dispersion relation
\begin{equation}
    \underset{L \to \infty}{\mathrm{lim}}E_{\rm SW}(k) = \frac{4\sqrt{2}}{3}~\mathrm{cos}\Bigl(\frac{k-\pi}{4}\Bigr)~\mathrm{sin} \Big(\frac{k}{2} \Bigr)^{5/2},
\end{equation}
with $k \in [0,2 \pi]$. The scaling of the gap follows from the energy of the first excited spin-wave state $E_{\rm SW}(k=2\pi/L) = \Delta_{\rm SW} \sim L^{-5/2}$.  Since the gap of the Fredkin Hamiltonian is upper-bounded as $\Delta \leq \Delta_{\rm SW}$, we obtain the following lower bound on the dynamical exponent:
\begin{equation}
    z \geq z_{\rm SW} = 5/2.
\end{equation}
The lower bound on the Fredkin dynamical exponent is our main result. To the best of our knowledge, this is the first proof that Fredkin-constrained systems exhibit subdiffusive transport; this follows from our physically-motivated spin-wave ansatz.  We emphasize that since we obtain the energy of the spin-wave state exactly, even for finite system size $L$ (see Appendix~\ref{section:bound}), this result is entirely exact.

\section{Discussion}


In this work, we identified a locally-conserved quantity that appears in Fredkin-constrained dynamics and used it to prove a subdiffusive lower bound for the Fredkin dynamical exponent; the proof of the bound is sufficient to establish that Fredkin-constrained dynamics belong to a dynamical universality class distinct from diffusion.  These conserved ``matched parentheses charges,'' which are irreducible Dyck words~\cite{Korepin2016fredkin}, have algebraically-decaying tails that underpin all non-trivial static correlations. Despite the fact that these charges are ``pseudolocal,'' they obey a local conservation law and are both extensive and additive, thus making them relevant to transport.  We construct a variational ansatz from the matched parentheses charge for the first excited state of the Fredkin spin chain in a fixed crystal momentum sector. This physically-motivated trial wavefunction provides an upper bound $\mathcal{O}(L^{-5/2})$ on the gap of the Fredkin spin chain, thus providing a lower bound on the dynamical exponent of Fredkin-constrained systems, i.e. $z \geq 5/2$. Our results may also be shown exactly for the Motzkin spin chain (Appendix~\ref{app:motzkin}).

The bound obtained from the spin-wave ansatz represents the first proof that transport in Fredkin-constrained systems is subdiffusive, which consequently implies that Fredkin-constrained dynamics belong to their own universality class distinct from ordinary diffusion.  While obtaining a subdiffusive bound constitutes significant progress in understanding Fredkin-constrained dynamics, the exact value of the dynamical exponent still remains an open question. A derivation of the true  dynamical exponent and detailing the mechanism responsible for its value would represent a clear direction for future research.  We also note that Fredkin and Motzkin constrained systems seem to be -- to the best of our knowledge -- one of the only examples of a system with local interactions but with a conserved charge having non-local densities as relevant hydrodynamic modes.  While the consequences of pseudolocal densities have mostly been explored for charges with exponentially decaying tails~\cite{Prosen_1998,PhysRevE.60.3949, PhysRevLett.106.217206, PhysRevLett.111.057203, doyon2017thermalization, doyon2022diffusion}, further exploring the consequences of densities with algebraically-decaying correlations would also be an interesting area for future research.





\section{Acknowledgements}

H.S. and R.V. thank Aaron Friedman and Brayden Ware for earlier collaborations on this topic. C.M. acknowledges support from NSF GRFP-1938059. This work was supported by the US Department of Energy, Office of Science, Basic Energy Sciences, under award No. DE-SC0023999 (H.S. and R.V.). S.G. and R.V. acknowledge hospitality of KITP during the DYNISQ22 follow-on program ``Phases of active quantum matter''. KITP is supported by grant NSF PHY-2309135.

\appendix

\section{Proof of additivity of matched parentheses charge}
\label{section:additivity}

In the main text, we argue that despite the fact that the matched parentheses charge has algebraically-decaying tails, the charge is extensive and additive and thus is sufficiently local to be relevant to thermodynamics and hydrodynamics.  In this section, we derive this statement explicitly. Consider a system of size $2L$ with open boundary conditions in a random (infinite temperature) state, with total number of matched parentheses $Q$. Let us cut the system into two halves $A$ and $B$ of size $L$, and denote by $Q_A$ and $Q_B$ the number of matched parentheses in subsystems $A$ and $B$, respectively. 

 In this appendix, we derive the full probability distribution of the charge $Q_A$ (or $Q_B$), and provide exact expressions for both $\langle Q_A \rangle$ and $\langle Q_A^2 \rangle_c$, showing that they scale asymptotically as $\sim L$. As a result, typical values of $Q_A$ and $Q_B$ are overwhelmingly likely to be equal to $L/2$, up to ${\cal O}(\sqrt{L})$ corrections. This implies that the  matched parentheses charge is additive: $Q \simeq Q_A + Q_B$, again up to subleading ${\cal O}(\sqrt{L})$ corrections. 

To establish this result, first consider a subsystem of $L$ sites in a fixed charge sector with $L-n$ open and $n$ closed parentheses.  The number of states that contain $Q$ matched parentheses charge is given by Catalan's triangle:
\begin{equation}
    T(L, Q) = \binom{L}{Q} - \binom{L}{Q-1},
\label{eqn:prob1}
\end{equation}
for $Q \leq \mathrm{min}(n, L-n)$.  
%
%
Generalizing to an infinite temperature state requires summing over all possible charge sectors.  Therefore, for a system of $L$ sites at infinite temperature, the number of possible states that have $Q_A$ matched parentheses is given by the following expression:
\begin{equation}
    T_{\rm inf}(L, Q_A) = (L-2Q_A+1) T(L, Q_A),
\label{eqn:prob2}
\end{equation}
for $Q_A \leq L/2$.   The probability $P_L( Q_A)$ of finding a state with $Q_A$ matched parentheses charges is thus given by:
\begin{equation}
    P_L(Q_A) = \frac{T_{\rm inf}(L, Q_A)}{2^L}.
\label{eqn:prob3}
\end{equation}
Given the probability of finding a subsystem of $L$ sites that contains $Q_A$ matched parentheses charges, it is straightforward to compute all moments of total matched parentheses charge $Q_A$ in that interval.  

%
%
%

We can derive the asymptotic behavior of these quantities in the limit $L \gg 1$ using Stirling's formula. Introducing the matched parentheses density $q_A=Q_A/L$, we find that the probability distribution $P_L(q_A) =P_L(Q_A=q_A L) L$ is maximized by the maximal allowed value of $q_A$, $q_A=q_\star=1/2$. Expanding about this maximum $q_A=1/2 - \delta$ with $\delta>0$, we find 
\begin{equation}
P_L(\delta)  \underset{L \to \infty}{\sim} 8 L^{3/2} \sqrt{\frac{2}{\pi}} \delta^2 {\rm e}^{- 2 L \delta^2}.
\end{equation}
A similar distribution in terms of height variables was derived in Refs.~\onlinecite{ChenFredkin2017, ChenFredkin2017a}. Using this distribution, we can derive the asymptotic behavior of all cumulants of $Q_A = L/2 - L \delta$. This yields 
\begin{equation}
 \langle Q_A \rangle = \frac{L}{2} - \sqrt{\frac{2L}{\pi}} + {\cal O}(1),
\end{equation}
and
\begin{equation}
 \langle Q_A^2 \rangle_c = \left(\frac{3}{4} - \frac{2}{\pi} \right) L + \dots
\end{equation}
As claimed in the main text, both the mean and variance of $Q_A$ scale linearly with the interval's size. In particular, fluctuations are supressed as $\sqrt{\langle Q_A^2 \rangle_c} / \langle Q_A \rangle \sim 1/\sqrt{L}$. This implies that the matched parentheses charge is additive: $Q \simeq Q_A + Q_B$, up to subleading ${\cal O}(\sqrt{L})$ corrections. 

In the above setup, the matched parentheses charge only receives contributions from matched parentheses residing within the region. One could also consider situations where the matched parentheses charge receive contributions outside the subregion. For instance, if one considers a subregion $[-\ell/2,\ell/2]$ of a larger system of size $L\gg \ell$ then any matched parentheses with ends located at $-y$ and $y$ (including $y>\ell/2$) contributes to the matched parentheses charge in that interval. Despite this difference, we have numerically confirmed that this does not change the scaling behavior of the matched parentheses charge with subregion size.

\section{Derivation of exact bound on gap scaling} \label{section:bound}

In the main text, we construct a parameter-free variational spin-wave ansatz for the Fredkin Hamiltonian in terms of the matched parentheses charge.  Here, we derive the an exact expression for the energy $E_{\rm SW}(k)$ of these spin-wave states, and use them to upper bound the gap of the Fredkin Hamiltonian, $\Delta \sim L^{-z}$, with $L$ the size of the system.  We emphasize that all of the expressions shown below are obtained exactly for finite-size systems.

\subsection{Spin wave ansatz and charge-current correlations}
\label{subsection:ansatz}

We construct a (unnormalized) spin-wave ansatz in terms of the the matched parentheses charge as $\ket{k} = \sum_n e^{i k \cdot n} q_n \ket{GS}$. Recall that $q_n = \sum_\ell q_n^\ell$ counts the total number of matched parentheses centered on link $n$, while $q_n^\ell = (a_n^\ell)^\dag a_n^\ell$ corresponds to a set of matched parentheses separated by $2\ell$ centered on link $n$.  As in the main text, we adopt the convention that lower indices denote position while upper indices denote separation; if either the upper or lower index is irrelevant to the discussion at hand, it will be omitted.  Here, we work in the $S^z_{\rm tot}=0$ ensemble, where the ground state of the Fredkin chain $ \ket{GS}$  is given by the uniform superposition of all $S^z_{\rm tot}=0$ states. 

Since $\ket{k}$ is a state with definite crystal momentum, the energy expectation value
\begin{equation}
E_{\rm SW}(k) = \frac{\bra{k} H \ket{k}}{\braket{k|k}} =  \frac{\sum_{n, m} e^{ik\cdot(n-m)} \bra{GS} q_m H q_n \ket{GS}}{\sum_{n, m} e^{ik\cdot(n-m)} \bra{GS} q_m q_n \ket{GS}},
\end{equation}
provides an upper bound on the ground state energy of the Hamiltonian $H$ in the sector with momentum $k$. Since the lowest excited-state of the Fredkin Hamiltonian is in the $k = 2 \pi/L$ sector, and since $H \ket{GS} = 0$, the variational spin-wave energy provides an upper bound on the Fredkin gap $\Delta =E(k= 2 \pi /L)  \leq E_{\rm SW}(k= 2 \pi /L)$. Because the total number of matched parentheses $\sum_n q_n$ is conserved, we have $E_{\rm SW}(k= 0)=0$, and so the spin-wave ansatz is guaranteed to be gapless as $k \to 0$.

Next, we invoke translation invariance and consider the quantity $\bra{GS} q_0 H q_n \ket{GS}$.  Using $H\ket{GS}=0$, this correlator can be expressed as $\bra{GS} q_0 [H, q_n] \ket{GS}$.  We note that $\dot q_n = i[H, q_n] = j_{n}-j_{n+1}$, where $j_n$ is the current of the matched parentheses charge on site $n$; therefore, this quantity can be expressed as a discrete derivative: 
\begin{equation}
    \bra{GS} q_0 H q_n \ket{GS} = i (  \langle q_0 j_{n+1} \rangle_0 -  \langle q_0 j_{n} \rangle_0),
\end{equation}
where we adopt the notation that $\langle \bullet \rangle_0$ corresponds to $\braket{GS| \dots |GS}$.  Therefore, obtaining an expression for the charge-current correlations $\langle q_0 j_n \rangle_0$ will directly lead to an exact bound for the scaling of $E(k=2\pi/L)$.  

\subsection{Source of charge-current correlations}
\label{subsection:correlations}

In the picture of the matched parentheses charge, it becomes clear that the source of all Fredkin dynamics is the movement of charges of semilength $\ell=1$, which are matched parentheses of the form (). When an () charge moves, it necessarily disrupts the end parenthesis of another charge of semilength $\ell$.  This can be pictured as the () moving past one of the matched end parentheses of the charge of semilength $\ell$, which we will call moving ``inside'' or ``outside'' of the charge of semilength $\ell$.  The movement of the () charge  triggers the process in which the charge of semilength $\ell$ hops by a single site to link $n=\pm 1$ while changing semilength by one to $\ell \pm 1$.  

All of these considerations are captured by the current operator $j_n = \sum_\ell j_n^\ell$, where each $j_n^\ell$ may be written explicitly in terms of ladder operators as:
\begin{equation}
    \begin{aligned}
        j_n^\ell = -\frac{i}{2} &\Bigl(~ (a_{n+\ell-1}^{\ell+1})^{\dagger}(a_n^1)^{\dagger}a_{n+\ell}^\ell a_{n-1}^1 - \\
        &- (a_{n-1}^{\ell+1})^{\dagger}(a_{n-\ell}^1)^{\dagger}a_n^\ell a_{n-\ell-1}^1  + \\
        &  +  (a_n^{\ell+1})^{\dagger}(a_{n+\ell-1}^1)^{\dagger}a_{n-1}^\ell a_{n+\ell}^1  - \\
        &- (a_{n-\ell}^{\ell+1})^{\dagger} (a_{n-1}^1)^{\dagger}a_{n-\ell-1}^\ell a_n^1  -\text{h.c.} ~\Bigr)
    \end{aligned}
\end{equation}
Since the only source of dynamics from the Fredkin Hamiltonian involve locally moving () charges, all nontrivial correlations between a charge sitting on link 0 and a current on site $n$ must come from the coincidence of two scenarios: first, that the charge sitting on link 0 corresponds to matched parentheses separated by $\ell \pm 1$, and second, that a () charge on link $n= \pm (\ell\mp 1)$ disturbs the end parenthesis of the charge on link 0.  This statement has an intuitive interpretation-- a current may only be correlated with a charge of semilength $\ell$ if a () charge either moves inside or outside the charge, which consequently changes the separation of the charge to $\ell \pm 1$.  In order to move inside or outside of the charge of semilength $\ell$ at a given timestep, since there are no non-local () hops encoded in the Fredkin Hamiltonian, there must be a () charge located directly adjacent to one of the matched end parentheses of the charge of semilength $\ell$ immediately before the hop.  Therefore, the task of understanding the static charge-current correlations in Fredkin reduces to finding the probability of these configurations.  Exact expressions for these probabilities are derived in the following section.

\subsection{Probabilities}

First, we focus on the process $()(\dots)^{\ell} \rightarrow (()\dots)^{\ell+1}$, in which $(\dots)^\ell$ pictorially represents some matched parentheses charge of semilenght $\ell$.  We denote the probability of finding a configuration of the form $()(\dots)^\ell$ or $(\dots)^\ell()$ conditional on having a $(\dots)^\ell$ as $p_{\rm out}(\ell)$. From direct counting, we find that $p_{\rm out}(\ell)=7/16$ in the thermodynamic limit since the system is locally at infinite temperature.  For a system of $L$ sites in the $S^z_{\rm tot}=0$ ensemble, however, the expression for $p_{\rm out}(\ell, L)$ becomes dependent on both $\ell$ and $L$ due to additional correlations that arise from finite-size effects.  

Since any Dyck word contains equal numbers of up and down spins, the $L-2\ell$ sites in the system that are not part of the $(\dots)_\ell$ charge must also contain the same number of up and down spins when the system is at $S^z_{\rm tot}=0$.  There are $\binom{L-2\ell}{(L-2\ell)/2}$ possible states for these $L-2\ell$ sites.  Consider a Dyck word of semilength $\ell+1$ of the form $()(\dots)^\ell$; since $()(\dots)^\ell$ also contains only matched sets of parentheses, the $L-(\ell+2)$ sites in the system that are not part of $q^\ell$ must contain the same number of up and down spins.  Therefore, there are $\binom{L-2\ell-2}{(L-2\ell-2)/2}$  possible configurations for these $L-2\ell-2$ sites. Finally, taking care to avoid double-counting the $\binom{L-2\ell-4}{(L-2\ell-4)/2}$ configurations of the form $()(\dots)^\ell()$, we obtain the following expression for $p_{\rm out}(\ell, L)$:
\begin{equation}
    p_{\rm out}(\ell, L) = \frac{2\binom{L-2\ell-2}{(L-2\ell-2)/2} -\binom{L-2\ell-4}{(L-2\ell-4)/2}}{\binom{L-2\ell}{(L-2\ell)/2}}.
    \label{eq:pout}
\end{equation}
Next, we focus on the process $(()\dots)^{\ell} \rightarrow ()(\dots)^{\ell-1}$.  We denote the probability of finding a configuration of the form $(() \dots)^\ell$ conditional on having a  $(\dots)^\ell$ as $p_{\rm in}(\ell)$.   Unlike when a charge of semilength $\ell=1$ sits directly outside of a charge of semilength $ell$, a charge of semilength $\ell=1$ sitting directly inside of a $q^\ell$ charge is subject to strong correlations from being embedded within a Dyck word.   In total, there are $C_{\ell-1}$ possible Dyck words of semilength $\ell$, where $C_\ell$ are Catalan numbers. Recall that this was obtained from there fact that of the $C_{\ell}$ Dyck word of semilength $\ell$, only those of the form $(\dots)^\ell$ correspond to matched parentheses charges; therefore, the $\dots$ in $(\dots)^\ell$ must also be a Dyck word of semilength $\ell-1$, of which there are $C_{\ell-1}$ possibilities.  By similar logic, there are $\ell-2$ ways to obtain a charge of the form $(()\dots)^\ell$, where $\dots$ is a path of length $\ell-2$.  To obtain the expression for $p_{\rm in}(\ell)$, we consider both words of the form $(()\dots)^\ell$ and $(\dots())^\ell$, of which there are $2~C_{\ell-2}$; however, care must be taken to not count configurations like $(()\dots())^\ell$ twice.  Since $\dots$ in these configurations must be a Dyck path of semilength $\ell-3$, the expression for $p_{\rm in}(\ell)$ is given by the following:
\begin{equation}
    p_{\rm in}(\ell) = \frac{2C_{\ell-2} - C_{\ell-3}}{C_{\ell-1}}.
    \label{eq:pin}
\end{equation}
We note that the boundary cases for $\ell \leq 2$ must be handled separately.  Since the only possible configuration for a $q^2$ charge is (()), $p_{\rm in}(2) = 1$.  Additionally, since there is no possible way for a $q^1$ charge to be embedded in another $q^1$ charge, $p_{\rm in}(1)=0$.  We also note that in the limit $\ell \gg 1$, the correlations that come from being embedded in a Dyck path become weaker and $p_{\rm in}(\ell) \rightarrow 7/16$. 

The expressions in eqs.~\ref{eq:pout} and~\ref{eq:pin} for $p_{\rm in}(\ell)$ and $p_{\rm out}(\ell, L)$ are conditioned on finding a charge of semilength $\ell$ on a particular link; however, in order to obtain the expression for $\langle q_0 j_n \rangle$, these conditional probabilities need to be multiplied by the probability to have a matched parentheses charge of semilength $\ell$ on link $n=0$ in the first place.  To produce the relevant expressions, we write down the probability $p_{q}(\ell, L)$ of finding a charge of semilength $\ell$ on any particular link in a system of $L$ sites.  Overall, there are $\binom{L}{L/2}$ possible states that the system can be found in as a while.  Since there are $C_{\ell-1}$ possible Dyck paths corresponding to a charge of semilength $\ell$, and since there are $\binom{L-2\ell}{(L-2\ell)/2}$ possible states that the the $L-2\ell$ sites outside of $(\dots)_\ell$ may take, $p_q(\ell, L)$ is given by the following expression:
 \begin{equation}
     p_{q}(\ell, L) = \frac{C_{\ell-1} ~\binom{L-2\ell}{(L-2\ell)/2}}{\binom{L}{L/2}}.
 \end{equation}

\subsection{Exact expression for $E_{\rm SW}(k)$}

With the probabilities $p_{\rm in}(\ell)$, $p_{\rm out}(\ell, L)$, and $p_q(\ell, L)$ obtained in the previous section, the charge-current correlator is given exactly by:
\begin{equation}
    \begin{aligned}
    i \langle q_0 j_n \rangle_0 = \frac{1}{2} ~ \Big( &p_{\rm out}(n-1, L) p_{q}(n-1, L) - \\ 
    &- p_{\rm in}(n+1) p_{q}(n+1, L) \Big).
    \end{aligned}
    \label{eq:fredkin_qj}
\end{equation}
From this, we may exactly compute the correlator $\bra{GS} q_0 H q_n \ket{GS} = i (  \langle q_0 j_{n+1} \rangle_0 -  \langle q_0 j_{n} \rangle_0)$ explicitly in terms of probabilities:
\begin{equation}
    \begin{aligned}
    \bra{GS}& q_0 H q_n \ket{GS} = \frac{i}{2}\big(
    ( p_{\rm out}(n,L)p_q(n,L) - \\ &- p_{\rm in}(n+2)p_q(n+2,L) ) - (p_{\rm out}(n-1,L) \\&p_q(n-1,L) - p_{\rm in}(n+1) p_q(n+1,L) )~\big)
    \end{aligned}
    \label{eq:corr_probs}
\end{equation}
The exact dispersion for the spin-wave state is given by:
\begin{equation}
E_{\rm SW}(k) = \frac{\sum_{n} ie^{ik\cdot n} (  \langle q_0 j_{n+1} \rangle_0 -  \langle q_0 j_{n} \rangle_0)}{(3L-4)/4(L-1)},
\end{equation}
where we used the normalization $\braket{k|k} = \frac{L( 3L - 4)}{ 4 ( L-1 )} $ valid for $k=2 \pi j/L$ with $j$ odd -- for $j>0$ even, this factor should be replaced by $\braket{k|k} =3(L/2)^2/(L-1)$.  This expression may used to evaluated the exact value of $E_{\rm SW}(k=2\pi/L)$ for any $L$.

The asymptotics of this expression may be derived in a straightforward way. In the thermodynamic limit $L \to \infty$, the correlator $\bra{GS} q_0 H q_x \ket{GS}$ decays as 
\begin{equation}
\bra{GS} q_0 H q_x \ket{GS} \underset{x \to \infty}{\sim} \frac{15}{64 \sqrt{\pi} x^{7/2}},
\end{equation}
at long distances, giving rise to a contribution $\sim k^{5/2}/(3 \sqrt{2}) $ to $E_{\rm SW}(k)$ at small momenta. In general, one would also expect a $\sim k^2$ contribution in $E_{\rm SW}(k)$ dominating this anomalous behavior coming from short distance physics. However, this diffusive contribution vanishes exactly because of the sum rule $\sum_n n^2 \bra{GS} q_0 H q_n \ket{GS} = 0$, which follows from our exact expressions. In the thermodynamic limit where $k$ becomes continuous, we find the exact dispersion relation
\begin{equation}
\underset{L \rightarrow \infty}{\rm lim} E_{\mathrm{SW}}(k) = \frac{4 \sqrt{2}}{3}  ~\mathrm{cos} \left(\frac{k-\pi}{4} \right)~\mathrm{sin} \left(\frac{k}{2} \right)^{5/2},
\end{equation}
valid when $k \in [0,2 \pi]$.
We thus conclude that $E_{\rm SW}(k) \sim k^{5/2}/(3 \sqrt{2}) $ as $k \to 0$, leading to the scaling $E_{\rm SW}(k = 2 \pi/L) = \Delta_{\rm SW}(L) \sim L^{-5/2}$ as $L \to \infty$.  This expression provides a rigorous upper bound on the Fredkin gap $\Delta = E(k = 2 \pi/L) $ .  In turn, this upper bound on the gap $\Delta \sim L^{-z}$ translates to a lower bound on the dynamical exponent
\begin{equation}
z \geq z_{\rm SW} = \frac{5}{2},
\label{eqn:bound}
\end{equation}
which is the main result of this letter.

\section{Numerical results for the Fredkin matched parentheses charge} \label{section:numerics}

\begin{figure*}[ht!]
    \centering
    \includegraphics[width=0.95\textwidth]{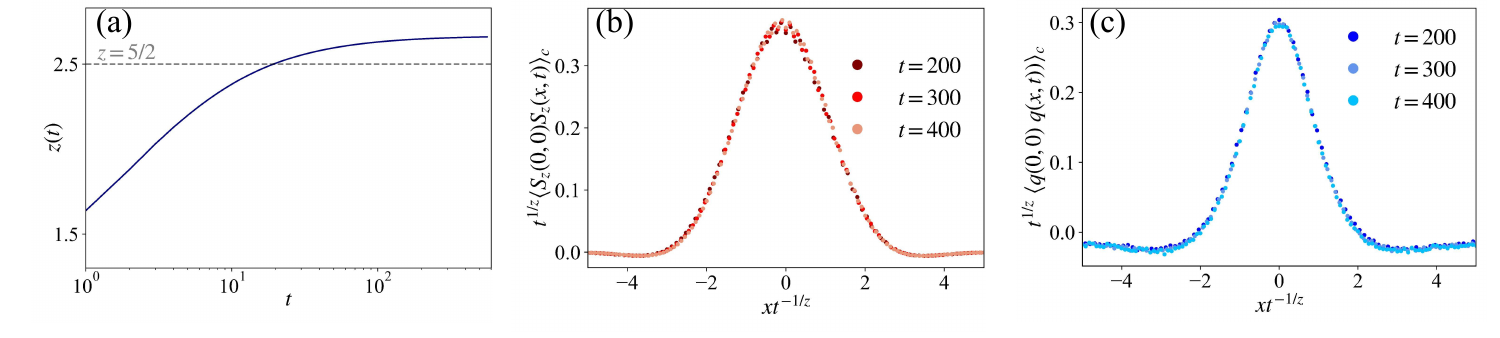}
    \caption{\textbf{Numerical results} (a) The dynamical exponent $z$ extracted from structure factor data for the Fredkin Markov process. At long times, $z$ appears to saturate to a value slightly larger that 5/2, although we cannot rule out logarithmic corrections.  The structure factor collapses for the original $U(1)$ charge (b) and the matched parentheses charge (c), using $z=5/2$, shown for times $t = 200-400$.  For all numerics, system sizes were chosen such that finite-size effects are negligible on the timescales shown.}
    \label{fig:supp}
\end{figure*}

Here, we briefly present some numerical results for the matched parentheses charge.  These numerics were run using the standard Markov process simulation techniques discussed in ref. \cite{hans_subdiffusion}.  As found in prior numerical studies on Fredkin-constrained dynamics, the numerically-extracted dynamical exponent appears to saturate to a value slightly larger than 5/2 (fig. \ref{fig:supp}a).  It is unclear from the numerics what the true value of the dynamical exponent is, as it is possible that the curve shown in fig. \ref{fig:supp}a could continue to slowly increase for arbitrarily long times; furthermore, the data is also consistent with $z=5/2$ with a lograrithmic correction, as suggested in ref. \cite{Koduvely1998}.  For numerically-accessible timescales, however, the structure factor for the $U(1)$ charge conserved by the Fredkin Markov process collapses neatly for $z \approx 5/2$ (fig. \ref{fig:supp}b).

In the main text, it was argued that the matched parentheses charge should have the same dynamical exponent as the original $U(1)$ charge conserved by the Markov process on account of the fact that each move of a matched parentheses charge corresponds to the movement of one of its endpoints (or both, in the case of the $\ell=1$ charge).  This intuition can be confirmed numerically; the structure factor for the matched parentheses charge collapses neatly for $z \approx 5/2$ (fig. \ref{fig:supp}c).

\section{Results for Motzkin}
\label{app:motzkin}

The main text and the previous two sections have dealt with Fredkin spin chains and Fredkin-constrained systems.  The Fredkin spin chain was originally proposed as the half-integer spin version of a model known as the Motzkin spin chain \cite{PhysRevLett.109.207202}.  Given the similarities between the two models, it should come as no surprise that the same procedure as the one used for Fredkin may be employed to derive an upper bound on the scaling of the gap of the Motzkin Hamiltonian, which also serves as a lower bound on the Motzkin dynamical exponent.

Here, mirroring the format of the main text, we outline the procedure to bound the gap of the Motzkin Hamiltonian.  After reviewing the Motzkin spin chain, we provide an overview of the relevant combinatorics and identify the spin-1 version of the matched parentheses charge.  As with Fredkin, the Motzkin matched parentheses charge can be shown to be pseudolocal in that, despite having power-law tails, it is extensive and additive.  The gap of the Motzkin Hamiltonian $\Delta_{\rm M}$ may then be bounded through the construction of a variational spin-wave ansatz, which features the same scaling as the gap of the Fredkin Hamiltonian, $\Delta_{\rm M} \leq {\cal O}(L^{-5/2})$.  We then arrive at our main result of a bound on the Motzkin dynamical exponent, $z \geq 5/2$.

\subsection{Background}

\subsubsection{The Motzkin spin chain}

The Motzkin spin chain was originally proposed as an interesting example of a frustration-free spin-1 chain with a highly-entangled ground state \cite{PhysRevLett.109.207202}.  With periodic boundary conditions, Motzkin Hamiltonain may be expressed as the sum of nearest-neighbor projectors, $H = \sum_{n, n+1} \Pi_{n, n+1}$.  Adopting the notation used by Ref.~\cite{ChenFredkin2017a}, the onsite Hilbert space is spanned by the states $\ket{u}$, $\ket{d}$ and $\ket{0}$, and the 2-site projector $\Pi_{n, n+1}$ may be expressed as:
\begin{equation}
    \Pi_{n, n+1} = \ket{D}_{n, n+1}\bra{D} + \ket{U}_{n, n+1}\bra{U} + \ket{V}_{n, n+1}\bra{V}, 
\end{equation}
where:
\begin{equation}
    \begin{aligned}
        &\ket{D} = \frac{1}{\sqrt{2}} ( \ket{0d} - \ket{d0} ), \\
        &\ket{U} = \frac{1}{\sqrt{2}} ( \ket{0u} - \ket{u0} ), \\
        &\ket{V} = \frac{1}{\sqrt{2}} ( \ket{00} - \ket{ud} ).
    \end{aligned}
\end{equation}
With open boundary conditions, the ground state of the Motzkin Hamiltonian for a system of $L$ sites is given by the uniform superposition of all Motzkin words of length $L$  (see the following subsection) \cite{PhysRevLett.109.207202}.  When considering periodic boundary conditions, the Motzkin ground state is instead given by the uniform superposition of all $S^z_{\rm tot} = 0$ states consisting of $L$ sites.

\subsubsection{Motzkin words}

In the context of Fredkin-constrained systems, spin-$\frac{1}{2}$ systems may be recast in terms of the Dyck combinatorial alphabet by replacing up-spins with an open parenthesis and down-spins with a closed parentheses.  The description of spin-1 models requires generalizing the Dyck alphabet to include a symbol for zero-spins, which may be accomplished by replacing zero-spins with a ``flat'' symbol, $-$.  Strings of parentheses and flats can be interpreted as a combinatorial object known as a Motzkin path; as suggested by the model's name, this a natural combinatorial language for Motzkin-constrained systems.  We briefly review some important aspects of Motzkin combinatorics here.

Motzkin words are strings of parentheses and flats that, read from left to right, do not contain unmatched parentheses.  For example, the Motzkin words of length $\ell=3$ are $- - -$, $- ()$, $(-)$, and $()-$.  The number of Motzkin words of length $\ell$ is given by the $\ell^{\rm th}$ Motzkin number, $M_\ell = \sum_{k=0}^{\ell/2} \binom{\ell}{2 k} ~C_k$.  Note that in Fredkin, we identify words by their \textit{semilength}, given that Dyck words necessarily contain an even number of parentheses; in contrast, we identify Motzkin words by their \textit{length} since the presence of flats allows for words to be comprised of an odd number of symbols.

Just as irreducible Dyck words are crucial to the construction of the Fredkin matched parentheses charge, irreducible Motzkin words will play the same role for this model. Motzkin words of length $\ell \geq 2$ that are irreducible begin and end with a matched set of parentheses; for example, the only irreducible Motzkin word of $\ell=3$ is $(-)$.  Since irreducible Motzkin words must be of the form $( \dots )$, where $\dots$ itself is a Motzkin word, there are $M_{\ell-2}$ irreducible Motzkin words of length $\ell$.  For the case of $\ell=1$, a single flat $-$ is itself an irreducible word. 

\subsection{The matched parentheses charge in Motzkin}

As outlined in the main text, each matched parentheses charge in Fredkin corresponds to the midpoint of a single irreducible word.  The conservation of the number of irreducible words in Fredkin therefore follows trivially from the fact that number of open (and closed) parentheses is conserved within a fixed charge sector.  In contrast, the number of irreducible words is not conserved in Motzkin for a fixed charge sector due to the allowed move $- - ~\leftrightarrow~ ()$, in which two flats (irreducible $\ell=1$ words) merge into a single irreducible word of length $\ell=2$.

A matched parentheses charge given by the midpoints of irreducible words may still be constructed for Motzkin-constrained system; however, the midpoints of paths of length $\ell \geq 2$ must correspond to two charges, while flats ($\ell=1$) only carry a single charge.  This modification ensures the conservation of the total matched parentheses charge during the allowed move $- - ~\leftrightarrow~ ()$, since there are two charges on either side of the process.  We will continue to refer to these charges at the matched parentheses charges and label them as $q_n^\ell$, where $n$ gives the location of the underlying irreducible word's midpoint and $\ell$ denotes the word's length.  Note that a consequence of the existence of odd-length words is that charges are permitted to live on both sites and links; therefore, the index for the matched parentheses charge runs from $n=1$ to $2L$.  In the $S_{\rm tot}^z=0$ charge sector for a system of $L$ sites with periodic boundary conditions, the total Motzkin matched parentheses charge is $Q = \sum_n q_n = L$.  

In each model, the smallest irreducible word plays the role of a freely-moving object that facilitates the motion of larger words when it disrupts one of their endpoints.  The allowed Fredkin move $()_n(\dots)_{n+\ell}^\ell \rightarrow (()_{n+1} \dots)_{n+\ell-1}^{\ell+1}$ is equivalent to the allowed Motzkin move $-_n(\dots)_{n+\ell+1}^\ell \rightarrow (-_{n+2} \dots)_{n+\ell}^{\ell+1}$ in that it shows the process of the smallest irreducible word of the model passing the end parenthesis of the word $(\dots)$, which leads to the displacement of the charge(s) corresponding to the larger word.  The fact that Motzkin charges may live on both links and sites leads to some notable differences in labelling the midpoint; in particular, the flat involved in this process changes its position by two, while the charges corresponding to the larger word only move one space.  Note that the process $- - \rightarrow ()$ behaves differently, as the flats each move one space when forming the $\ell=2$ word; therefore, this process must be dealt with carefully and separately.

Taking into account all of the considerations outlined above, the Motzkin Hamiltonian may be rewritten in terms of the matched parentheses charge.  Writing the hard-core bosonic ladder operators for matched parentheses charges of length $\ell$ whose midpoint position is $i$ as $a_i^{\ell (\dag)}$, the Hamiltonian in the $S^z_{\rm tot}$ charge sector with PBCs is given by:
%

\begin{equation}
    \begin{aligned}
        H& = \frac{1}{4}~ \Bigg\{~ \frac{1}{2}\sum_{i=1}^{2L}\sum_{\ell=2}^{L-1} \bigg( -\Big( 
        (a_{i-1}^{\ell+1 \dagger})^2 a_{i-\ell+1}^{1 \dagger} (a_{i}^{\ell})^2 a_{i-\ell-1}^{1}+ \\
        &+(a_{i+1}^{\ell+1 \dagger})^2 a_{i+\ell-1}^{1 \dagger} (a_{i}^{\ell})^{2} a_{i+\ell+1}^{1} + \text{h.c.} \Big) +
        \\
        & +\Big( (a_{i}^{\ell \dagger})^2(a_{i}^{\ell})^2  n_{i-\ell-1}^{1}+
        (a_{i-1}^{\ell+1 \dagger})^2(a_{i-1}^{\ell+1})^2 n_{i-\ell+1}^{1} + \\ &+(a_{i}^{\ell \dagger})^2(a_{i}^{\ell})^2 n_{i+\ell+1}^{1}+ (a_{i+1}^{\ell+1\dagger})^2(a_{i+1}^{\ell+1})^2 n_{i+\ell-1}^{1} \Big) \bigg) -\\
        &-\sum_{i=1}^{2L}\bigg(((a_{i+1}^{2 \dagger})^2 a_{i}^{1}a_{i+2}^{1}+\text{h.c})- \frac{1}{2}(a_{i}^{2 \dagger})^2(a_{i}^{2})^2 \bigg)+ \\
        &+ 2\sum_{i=1}^{2L} n^{1}_i n^{1}_{i+2} \Bigg\}.
    \end{aligned}
\end{equation}

\subsection{Pseudolocality of the Motzkin matched parentheses charge}

Like the matched parentheses charge in Fredkin, the Motzkin matched parentheses charge is \textit{pseudolocal} in that it obeys a local conservation law despite possessing power-law tails.  Here, we first show that the Motzkin matched parentheses charge has algebraically-decaying tails before proving that the charge in Motzkin is both extensive and additive.  Both properties together ensure that this charge is a relevant for transport.

\subsubsection{Power-law tails}

The main text outlines how all static correlations in spin-$\frac{1}{2}$ systems may be traced back to the fact that each endpoint of an irreducible word uniquely belongs to a single irreducible word.  The exact same argument holds for spin-1 systems.  In a spin-1 system, the existence of a single flat on site 0 precludes the simultaneous existence of any charge of length $\ell$ on either a link or site $\pm \ell$ away from the flat.

The Motzkin numbers are known to asymptotically scale as $M_\ell \sim 3^\ell ( 3/\ell )^{3/2}$ as $\ell \rightarrow \infty$.  Since there are $M_{\ell-2}$ possible irreducible Motzkin words of length $\ell$ and $3^\ell$ possible $\ell$-site states, the probability of finding a matched parentheses charge on any given site (or link) is given asymptotically by $M_{\ell-2}/3^\ell \sim \ell^{-3/2}$.  Therefore, the decay of the tails of the Motzkin matched parentheses charge, like the Fredkin case, is captured by $\langle q_m^1 q_n \rangle \sim |m-n|^{-3/2}$.

\subsubsection{Extensivity and additivity}

Given the myriad of similarities between Fredkin and Motzkin, it is unsurprising that the Motzkin matched parentheses charge is also extensive and additive.  To see this, consider the same setup outlined in Section \ref{section:additivity} but for a spin-1 system.  By only performing allowed Motzkin moves within subsystem $A$, the subsystem may be reordered into a state where all of the $n_-$ flats come first, followed by all of the $n_) + n_($ parentheses.  Since this state may be reached by only performing allowed Motzkin moves, the total charge of the subsystem $Q_A = \sum_{n=1}^{2L} q_n = \big( \sum_{n=1}^{n_-} q_n \big) + \big(\sum_{n=n_- + 1}^{2L} q_n\big)$ remains unchanged after this reordering.  The number of charges contributed by the flats trivially scales with $n_-$, and the number of charges attached to parentheses scales with $n_) + n_($ by the arguments presented in section \ref{section:additivity}.  Therefore, $Q_A \sim n_- + n_) + n_( = L$, so the charge is naturally extensive.

To show that the charge is additive,we derive the probability distribution of the charge in a subsystem.  By counting the population of flats and matched parentheses separately, it can be shown that the number of states in subsystem $A$ that contain $Q_A$ charges is given by:
\begin{equation}
    T_{\rm M}(L, Q_A) = \sum_{i=\mathrm{mod}(Q_A,2)}^{Q_A} \binom{L}{i}~ T_{\rm inf}\Bigl(L-i,\frac{Q_A-i}{2}\Bigr),
    \label{eq:Tmotzkin}
\end{equation}
where $T_{\rm inf}(L,Q)$ is given by eq. \ref{eqn:prob2}.  Therefore, the probability finding $Q_A$ charges in subsystem $A$ is given by:
\begin{equation}
    P_{L}(Q_A) = \frac{T_{\rm M}(L,Q_A)}{3^L}.
\end{equation}
Computing cumulants from this distribution, we find that both the mean and variance scale linearly with $L$, as in the Fredkin case. 

\subsection{Bound for the Motzkin gap scaling}

We constructed a spin-wave ansatz in order to bound the scaling of the gap of the Fredkin Hamiltonian, $\Delta \sim L^{-z}$, with $z \geq 5/2$.  Here, we employ a nearly identical approach in order to obtain the same bound for the Motzkin Hamiltonian.

\subsubsection{The Motzkin spin-wave ansatz}

We construct a spin-wave ansatz in terms of the Motzkin matched parentheses charge as $\ket{k} = \sum_n e^{i k \cdot n/2} q_n \ket{GS}$; the factor of $1/2$ in the exponent is due to the fact that the matched parentheses charges may sit on both sites and links, and with our conventions $n$ runs from 1 to $2L$.  As before, we work in the $S^z_{\rm tot}$ ensemble, where the ground state $\ket{GS}$ is the uniform superposition of all $S^z_{\rm tot}$ states. The energy expectation value is therefore given by:
\begin{equation}
    \begin{aligned}
    E_{\rm SW}(k) &= \frac{\bra{k} H \ket{k}}{\braket{k|k}} = \\ &= \frac{\sum_{n, m} e^{ik\cdot(n-m)/2} \bra{GS} q_m H q_n \ket{GS}}{\sum_{n, m} e^{ik\cdot(n-m)/2} \bra{GS} q_m q_n \ket{GS}}.   
    \end{aligned}
\end{equation}
We will admit that as in the Fredkin case, the norm just scales linearly with system size $\braket{k|k} \sim L$, so we will focus on the numerator in the following. Using translation invariance, we thus have 
\begin{equation}
    \begin{aligned}
    E_{\rm SW}(k) &\sim \underbrace{\sum_{n} e^{ik\cdot n/2} \bra{GS} q_0 H q_n \ket{GS}}_{E^{0}_{\rm SW}(k)} + \\
    &+ \underbrace{\sum_{n} e^{ik\cdot (n-1)/2} \bra{GS} q_1 H q_n \ket{GS}}_{E^{1}_{\rm SW}(k)}.
    \end{aligned}
\label{eq:motzkinsw}
\end{equation}
In our conventions, $n$ even corresponds to links, while $n$ odd corresponds to sites. 

\subsubsection{Charge-current correlations}


For Motzkin, the lattice continuity equation for the new conserved charge reads $i[H,q_n] = j_n-j_{n+1} + j'_n - j'_{n+2} $ with $j_n = \sum_{n=1}^{L-1}j^{\ell}_n$ and $j'_n = \sum_{\ell=2}^{L-1}j_n^{'\ell}$.  The current operator $j_n^\ell$ associated with the movement of charges that are associated with matched sets of parentheses is given by:
\begin{equation}
j_{n}^{\ell} = \begin{cases}
-\frac{i}{4}\bigg(a_{n-1}^1 a_{n+1}^1 (a_n^{2 \dagger})^2-a_{n-2}^1 a_{n}^1 (a_{n-1}^{2 \dagger})^2-\text{h.c.}\bigg), \\ ~~~\text{if $\ell=1$,} \\ \\
 \frac{i}{4}\bigg( 
(a_{n-1}^{\ell+1 \dagger})^2 a_{n-\ell+1}^{1 \dagger} (a_{n}^{\ell})^2 a_{n-\ell-1}^1- \\ -(a_{n}^{\ell+1 \dagger})^2 a_{n+\ell-2}^{1 \dagger} (a_{n-1}^{\ell})^2 a_{n+\ell}^1-\text{h.c.} \bigg), \\ \text{otherwise,}
\end{cases}
\end{equation}
while the current operator $j_n^{' \ell}$ corresponding the movement of flats is given by:
\begin{equation}
    \begin{aligned}
    j_{n}^{' \ell} = -\frac{i}{8}\bigg( (a_{n+\ell-2}^{\ell+1 \dagger})^2 a_{n}^{1 \dagger}(a_{n+\ell-1}^{\ell})^2 a_{n-2}^1 - \\ - (a_{n-\ell}^{\ell+1 \dagger})^2 a_{n-2}^{1 \dagger} (a_{n-\ell-1}^{\ell})^2 a_{n}^{1}-\text{h.c.}\bigg).
    \end{aligned}
\end{equation}
The origin of charge-current correlations may be understood in direct analogy to the source of charge-current correlations in Fredkin, as outlined in \ref{subsection:correlations}.  In Motzkin, the movement of any charge of length $\ell > 2$ with position $i$ is facilitated by a single flat passing through one of its endpoints.  As the flat disrupts one of the endpoints of the larger charge, the larger charge changes in both length to $\ell \pm 1$ and position to ${i \pm 1}$. 

As already noted, the continuity equation contains contributions from two different types of currents of the matched parentheses charges ($j$ and $j'$) due to differences between the movement of flats and charges with length $\ell \geq 2$. During the process of passing through the endpoint of a larger word, the flat changes position by \textit{two}, since the flat may only live on sites.  In contrast, the other charge involved in this process will move by only \textit{one} space as it increases or decreases in length by one.  Therefore, since $\ell=1$ charges (flats) move different amounts than $\ell \geq 2$ charges, the continuity equation associated with the movement of $q_n$ cares whether the matched parentheses current at $n$ is caused by a $\ell=1$ or $\ell \geq 1$ charges.

We may follow the procedure used in \ref{subsection:ansatz} and write $\braket{GS| q_m H q_n | GS} = \braket{GS| q_m [H, q_n]| GS}$.  The energy of the spin-wave state may then be expressed as a sum of terms of the form $\langle q_m j_n' \rangle$ and $\langle q_m j_n \rangle$.  As was the case for Fredkin, these charge-current correlators may be expressed in terms of the probability of finding specific configurations-- we derive these probabilities in the following section.

\subsubsection{Probabilities}

In subsection \ref{subsection:correlations}, we outline how all charge-current correlations in Fredkin may be linked to the probability of finding configurations of the form $()(\dots)$ or $(() \dots)$, since the $\ell=1$ Dyck word facilitates the motion of larger words by passing through their endpoints.  In Motzkin, the same type of reasoning holds, except the $\ell=1$ Motzkin word is a single flat, $-$, instead of the $\ell=1$ Dyck word, ().  Therefore, charge-current correlations may be expressed in terms of the probability that configurations of the form $-(\dots)$ or $(- \dots)$ occur.

As was the case in Fredkin, the probability of finding a configuration like $(- \dots)$ is independent of system size.  There are $M(\ell)$ Motzkin paths of length $\ell$, $M(\ell -2)$ of which are irreducible.  For Fredkin, eq. \ref{eq:pin} gives the probability that there is an $\ell=1$ charge sitting directly inside of \textit{either} of the endpoints of the $\ell > 1$ charge; for reasons that will become clear in the following section, we are interested in looking at only \textit{one} of the endpoints for Motzkin.  Therefore, conditional on having a word of length $\ell$ on a particular link, we have:
%
%
%
\begin{equation}
    p_{\rm in}(\ell) = \frac{M(\ell-3)}{M(\ell - 2)}.
\end{equation}
This expression approaches $5/9$ in the $\ell \to \infty$ limit.  

Directly counting all possible configurations gives $p_{\rm out} = 5/9$ but does not account for finite-size effects.  Since there are $\frac{L!}{(n!)^2(L-2n)!}$ possible $S^z_{\rm tot}=0$ states with $L-2n$ spins-0, there are:
\begin{equation}
    \Omega(L) = \sum_{n=0}^{L/2} ~\frac{L!}{(n!)^2(L-2n)!},
\end{equation}
total possible states.  Therefore, the probability of finding a configuration of the form $- (\dots)$ is given by:
%
%
\begin{equation}
    p_{\rm out}(\ell, L) = \frac{\Omega(L-\ell-1)}{\Omega(L - \ell)}.
\end{equation}
Both $p_{\rm in}(\ell)$ and $p_{\rm out}(\ell, L)$ given above are conditioned on having a Motzkin word on a particular link; obtaining the full probability requires removing that condition.  The probability that any given link hosts a Motzkin word of length $\ell$ is given by:
\begin{equation}
    p_q(\ell, L) = \frac{M(\ell - 2) \Omega(L-\ell)}{\Omega(L)}.
\end{equation}

Having worked out all relevant probabilities, we can now use them to exactly find $E_{\rm SW}(k)$ and thus bound the scaling of the gap of the Motzkin Hamiltonian. 

\subsubsection{Motzkin spin-wave energy $E_{\rm SW}(k)$}

The energy of the spin-wave state splits into two parts, $E_{\rm SW} = E_{\rm SW}^0(k) + E_{\rm SW}^1(k)$, as shown in \ref{eq:motzkinsw}.   Each $E_{\rm SW}^m(k)$ has contributions from both $j_n$ currents and $j'_n$ currents:
\begin{equation}
    \begin{aligned}
    E_{\rm SW}^{m=0,1}(k) &\sim  \sum_{n} ie^{ik\cdot n/2} \Big( \big(  \langle q_m j_{n+1} \rangle_0 -  \langle q_m j_{n} \rangle_0 \big) +  \\ &+ \big( \langle q_m j_{n+2}' \rangle_0 - \langle q_m j_{n}' \rangle_0 \big) \Big), 
    \end{aligned}
\end{equation}
where again $\langle \bullet \rangle_0$ corresponds to $\braket{GS| \dots |GS}$. The form of the $\langle q_m j'_n \rangle$ depends on whether $m$ is even or odd.  For all values of $n$ and $m=0$, we have:
\begin{equation}
    \centering
    i \langle q_m j_n' \rangle_0 = \begin{cases}
        p_{\rm out}(n-1, L) p_{q}(n-1, L) - \\ ~~-
        p_{\rm in}(n+1) p_{q}(n+1, L) , \text{ for odd} ~n; \\ \\
        0 ~\mathrm{otherwise},
    \end{cases}
    \label{eq:motzkinqj'}
\end{equation}
while for $m=1$ and $n >3$, we have:
\begin{equation}
    \centering
    i \langle q_m j_n' \rangle_0 = \begin{cases}
        p_{\rm out}(n-2, L) p_{q}(n-2, L) - \\ ~~-
        p_{\rm in}(n) p_{q}(n, L) , \text{ for odd} ~n; \\ \\
        0 ~\mathrm{otherwise},
    \end{cases}
    \label{eq:motzkinqj'}
\end{equation}
For small values of $n$, correlations are given by $i \langle q_1 j_1' \rangle_0 = (\Omega(L-1)-\Omega(L-2)+2\Omega(L-3))/(2 \Omega(L))$, and  $i \langle q_1 j_3' \rangle_0 = -i \langle q_1 j_1' \rangle_0$. We again have $i \langle q_1 j_0' \rangle_0 = 0$.

The interpretation of the $i \langle q_1 j_n' \rangle_0$ correlations 
is straightforward: since $j'_n$ is the current on link $n$ associated with a flat moving across that link, there can be no $j'_n$ current associated with even values of $n$.  These expressions for odd $n$ is similar to the charge-current correlations in Fredkin (eq. \ref{eq:fredkin_qj}).

The form of the $\langle q_m j_n \rangle_0$ correlator depends on whether $m$ is even or odd due to the fact that $j_n$ is the current associated with $\ell>1$ irreducible words at $n$.  Therefore, all charge-current correlations are driven by the presence of an $\ell=1$ charge sitting at site $m$.  For link $m=0$, the $j_n$ correlations vanish for $|n| \geq 2$ since flats can not live on links; for $|n|=1$ and $m=0$, all correlations are driven by the presence of a single flat on site $n$, and so $i \langle q_1 j_1 \rangle_0 = p_q(1, L)$.  In contrast, flats are allowed to sit on site $m=1$, and so there are nontrivial charge correlations for all values of $n$ when $m=1$.  Therefore, we have:
\begin{equation}
    \begin{aligned}
    i \langle q_1 j_n \rangle_0 &=
        p_{\rm out}(n-2, L) p_{q}(n-2, L) - \\ 
        &- p_{\rm in}(n+1) p_{q}(n+1, L),
    \end{aligned}
    \label{motzkinqj}
\end{equation}
for $n > 3$.  For small values of $n$, we have $i \langle q_1 j_3 \rangle_0 =(p_q(2, L) - p_q(4,L))/2$, with $i \langle q_1 j_0 \rangle_0 = -i \langle q_1 j_3 \rangle_0$, and $i \langle q_1 j_1' \rangle_0 = (\Omega(L-1)-\Omega(L-2)+2\Omega(L-3))/(2 \Omega(L))$, with $i \langle q_1 j_2' \rangle_0 = -i \langle q_1 j_2' \rangle_0$.

Note that both $E_{\rm SW}^0$ and $E_{\rm SW}^1$ contain a diffusive $\sim k^2$ term at low momenta, but those contributions come in with opposite signs and exactly cancel each other out in $E_{\rm SW}$. This cancellation is due to a sum rule $\sum_{m, n} (m-n)^2 \braket{GS| q_m H q_n| GS} = 0$, similar to the one we showed in the Fredkin case, which rules out a potentially diffusive $\sim k^2$ contribution in $E_{\rm SW}(k)$. Instead, the low-energy dispersion is governed by the long distance of the correlator~\eqref{eq:motzkinqj'}, in close analogy with the Fredkin case~\eqref{eq:fredkin_qj}. The probabilities $p_{\rm in}$ and $p_{\rm out}$ are ${\cal O}(1)$ at long distances, whereas $p_q(\ell) \sim \ell^{-3/2}$. The charge current correlator thus scales as $i \langle q_0 j_n' \rangle_0 \sim n^{-5/2}$, and $E_{\rm SW}(k) \sim k^{5/2}$, mirroring the Fredkin spin-wave dispersion relation. Setting $k=2\pi/L$, this provides an upper bound ${\cal O}(L^{-5/2})$ on the gap of the Motzkin spin chain.

\bibliographystyle{unsrt}
\bibliography{refs}

\end{document}